\documentclass[twocolumn]{aastex63}

\newcommand{\beq}{\begin{equation}}
\newcommand{\eeq}{\end{equation}}
\usepackage{amsmath}
\usepackage{longtable}

\newcommand{\AU}{{\rm AU}} 
\newcommand{\kB}{k_{\rm B}}

\submitjournal{ApJ}

\shorttitle{Ring formation by dust coagulation}
\shortauthors{Ohashi et al.}
\graphicspath{{./}{figures/}}
\begin{document}

\title{Ring formation by coagulation of dust aggregates in early phase of disk evolution around a protostar}

\author[0000-0002-9661-7958]{Satoshi Ohashi}
\affil{RIKEN Cluster for Pioneering Research, 2-1, Hirosawa, Wako-shi, Saitama 351-0198, Japan}
\email{satoshi.ohashi@riken.jp}

\author[0000-0001-8808-2132]{Hiroshi Kobayashi}
\affil{Department of Physics, Graduate School of Science, Nagoya University, Furo-cho, Chikusa-ku, Nagoya 464-8602, Japan}

\author[0000-0002-1803-0203]{Riouhei Nakatani}
\affil{RIKEN Cluster for Pioneering Research, 2-1, Hirosawa, Wako-shi, Saitama 351-0198, Japan}

\author[0000-0002-1886-0880]{Satoshi Okuzumi}
\affil{Department of Earth and Planetary Sciences, Tokyo Institute of Technology, Meguro-ku, Tokyo 152-8551, Japan}

\author[0000-0001-9659-658X]{Hidekazu Tanaka}
\affil{Astronomical Institute, Tohoku University, 6-3 Aramaki, Aoba-ku, Sendai 980-8578, Japan}

\author[0000-0001-9843-2909]{Koji Murakawa}
\affil{Institute of Education, Osaka Sangyo University, 3-1-1 Nakagaito, Daito, Osaka 574-8530, Japan}

\author[0000-0001-7511-0034]{Yichen Zhang}
\affil{RIKEN Cluster for Pioneering Research, 2-1, Hirosawa, Wako-shi, Saitama 351-0198, Japan}

\author[0000-0003-2300-2626]{Hauyu Baobab Liu}
\affil{Academia Sinica Institute of Astronomy and Astrophysics, P.O. Box 23-141, Taipei 10617, Taiwan}

\author[0000-0002-3297-4497]{Nami Sakai}
\affil{RIKEN Cluster for Pioneering Research, 2-1, Hirosawa, Wako-shi, Saitama 351-0198, Japan}

%\nocollaboration{2}

%% Note that the \and command from previous versions of AASTeX is now
%% depreciated in this version as it is no longer necessary. AASTeX 
%% automatically takes care of all commas and "and"s between authors names.

%% AASTeX 6.3 has the new \collaboration and \nocollaboration commands to
%% provide the collaboration status of a group of authors. These commands 
%% can be used either before or after the list of corresponding authors. The
%% argument for \collaboration is the collaboration identifier. Authors are
%% encouraged to surround collaboration identifiers with ()s. The 
%% \nocollaboration command takes no argument and exists to indicate that
%% the nearby authors are not part of surrounding collaborations.

%% Mark off the abstract in the ``abstract'' environment. 
\begin{abstract}

Ring structures are observed by (sub-)millimeter dust continuum emission in various circumstellar disks from early stages of Class 0 and I to late stage of Class II young stellar objects (YSOs).
In this paper, we study one of the possible scenarios of such ring formation in early stage, which is coagulation of dust aggregates.
The dust grains grow in an inside-out manner because the growth timescale is roughly proportional to the orbital period.
The boundary of the dust evolution can be regarded as the growth front, where the growth time is comparable to the disk age.
With radiative transfer calculations based on the dust coagulation model, we find that the growth front can be observed as a ring structure because dust surface density is sharply changed at this position. Furthermore, we confirm that the observed ring positions in the YSOs with an age of $\lesssim1$ Myr are consistent with the growth front.
The growth front could be important to create the ring structure in particular for early stage of the disk evolution such as Class 0 and I sources.

\end{abstract}

%% Keywords should appear after the \end{abstract} command. 
%% See the online documentation for the full list of available subject
%% keywords and the rules for their use.
\keywords{Protostars
---Circumstellar dust
---Protoplanetary disks
---Star formation}

%% From the front matter, we move on to the body of the paper.
%% Sections are demarcated by \section and \subsection, respectively.
%% Observe the use of the LaTeX \label
%% command after the \subsection to give a symbolic KEY to the
%% subsection for cross-referencing in a \ref command.
%% You can use LaTeX's \ref and \label commands to keep track of
%% cross-references to sections, equations, tables, and figures.
%% That way, if you change the order of any elements, LaTeX will
%% automatically renumber them.
%%
%% We recommend that authors also use the natbib \citep
%% and \citet commands to identify citations.  The citations are
%% tied to the reference list via symbolic KEYs. The KEY corresponds
%% to the KEY in the \bibitem in the reference list below. 

\section{Introduction} \label{sec:intro}

A Keplarian disk is formed around a protostar and plays an essential role in planet formation.
Recent high spatial resolution observations of dust continuum with interferometers uncover a variety of pictures of disk structures  from the early stage of the disk formation \citep[e.g.,][]{tak17,she17,she18,sai20,tob20,gar20} to the late stage of protoplanetary disks \citep[e.g.,][]{van13,cas13,alma15,ise16,and18,fed18,tsu19}.

In particular, dust ring structures have been observed in many protoplanetary disks.
For example, the first ALMA Long Baseline Campaign observations show the prominent ring and gap structures of the HL Tau disk at 30 mas resolution \citep{alma15}.
The Disk Substructures at High Angular Resolution Project (DSHARP)  project in the ALMA Cycle 4 Large program \citep{and18} has also shown that disks have gap/ring structures in the sample of large and bright disks \citep{hua18}.

The observed ring structures are believed to be caused by changes in the density or dust opacities over the disk.
The formation mechanism of such ring structures remains unknown even though various mechanisms to create dust rings are proposed, such as the density gap formed by the gravitational interaction between the disk and unseen giant planets \citep[e.g.,][]{gol80,nel00,paa04,zhu12,kan15,zha18}, magnetorotational instability (MRI) \citep{flo15}, magneto-hydrodynamics (MHD) wind \citep{rio19,sur19}, both snow line and non-MHD effects \citep{hu19}, secular gravitational instability \citep{tak14}, snow lines of molecules or dust sintering \citep{zha15,oku16}, and so on.
These studies are mainly motivated by the observations for the protoplanetary disks around Class II sources.

Interestingly, the dust ring structures are observed not only in class II protoplanetary disks, but also even in the earlier stages of disk formation around class 0 and class I objects \citep[e.g.,][]{she17,she18,she20,nak20}.
The formation of the ring structures in growing young disks require much shorter time and its mechanisms could be constrained more.
Such approach has been started after the discovery of infant disks around young embedded protostar \citep{tob12,oha14,yen14}.
For example, gap formation by (an) unseen planet(s) has a difficulty of forming planets at such an early stage.
A MHD wind is proposed to create a ring structure in such young disks.
\citet{tak18} showed that the MHD wind creates a hole structure in young disks by losing disk materials inside the MHD wind.

In this paper, we investigate an alternative scenario of the ring-formation mechanism in the early phase of disk evolution.
We focus on the evolutionary process of dust aggregates by coagulation because grain growth is important in the disks.
The evolution of dust coagulation model has been studied by \citet{nak81,tan05,dul05,bra08,bir10,oku12} and others in order to investigate the formation of planetesimals or planets, but its connection to the ring structures have not been explored.
By using this dust model, we discuss whether the dust evolution by the coagulation can be observed as a ring structure during the grain growth.

\section{A picture of the Keplerian disk formation with dust evolution}

Before investigating the dust growth by coagulation in a Kepler disk, we mention the dust evolution process from the infalling envelope to the disk region.

Stars are formed via gravitational collapse in dense cores and protoplanetary disks are formed as byproducts of the star formation \citep[e.g.,][]{wil11}. 
A Keplarian-rotation disk is formed around a protostar in a slowly rotating dense core. The Kepler disk expands by the accretion of the infalling envelope.

\citet{hir09} calculated the dust coagulation in collapsing pre-stellar cores and showed that the dust grain does not grow in the envelope because the density of $10^{4-7}$ cm$^{-3}$ is not high enough to grow the dust grains.
\citet{orm09} also investigated the coagulation in dense cores and suggested that the free fall time of the collapsing cores is  not long enough to proceed the grain growth during the gravitationally collapsing. 
Therefore, we assume the dust coagulation in the Kepler disk region rather than the infalling envelope.
The idea of the ring formation by dust coagulation works as far as there is a Keplerian disk even though the disk is embedded in the envelope because the infalling materials will accrete to the outer edge of the disk, while the ring formation is considered inside the disk.
 
The Kepler disk becomes larger as accreting materials in the infalling envelope. 
The disk growth may need to be took into account with the dust growth.
However, this study assumes that the disk is already formed and then the dust coagulation occurs.
This is a case that the timescale of the disk evolution is faster than that of the dust coagulation in the disk.
We note that further studies are needed to investigate whether the disk growth is earlier than the dust growth.

\section{Theoretical Model of Dust Growth} \label{sec:theory}

\subsection{Disk Model}

In this subsection, we describe a dust coagulation model.
The results of the evolution of the surface density and grain size are shown in section \ref{evolution}.

A simple method for calculating dust surface density and particle size distribution is adopted according to \citet{sat16}, who investigated the water composition of planets
due to ice pebble accretion across the snow line. 
We consider two factors of dust evolution: (1) the grain growth via coagulation and (2) radial drift of dust grains.

The initial gas surface density ($\Sigma_{\rm g}$) is set to be 
\beq
{\Sigma}_{\rm g} = 1.7\times10^3\left(\frac{r}{1\ \AU}\right)^{-3/2} {\rm g~cm^{-2}}, 
\label{eq:Sigmag}
\eeq
based on the minimum mass solar nebula (MMSN) model of \citet{hay81}.
We ignore the jump of the surface density due to the snow line because the position of the snow line is $\sim3$ au which is smaller than the area to be investigated and observed ring positions ($\sim10-100$ au).
The initial dust surface density ($\Sigma_{\rm d}$) is set to be $1\%$ of the dust surface density.

The dust temperature of the disk is determined by assuming a thermal equilibrium as follows
\beq
{T} = 280\left(\frac{r}{1\ \AU}\right)^{-1/2} \left(\frac{L}{L_\odot}\right)^{1/4}{\rm K}.
\label{eq:Sigmag}
\eeq

The initial size of the dust particle is uniformly set to 0.1 $\mu$m.

The mass distribution of dust particles is assumed to have a single peak in mass $m_{\rm p}(r)$ at each radial distance $r$. 
Then, assuming that the dust surface density $\Sigma_{\rm d}$ at each orbit $r$ is dominated by particles with mass $m_{\rm p}$, we follow how the peaks of dust surface density $\Sigma_{\rm d}$ and mass $m_{\rm p}$ change with coagulation and radial drift.
Note that we assume that the dust aggregates are so sticky that no fragmentation or bouncing occurs upon collision.
According to \citet{sat16}, the equations of the evolution of $\Sigma_{\rm d}$ and $m_{\rm p}$ are given by 
\begin{equation}
\frac{\partial \Sigma_{\rm d}}{\partial t}+\frac{1}{r}\frac{\partial}{\partial r}(rv_{\rm r}\Sigma_{\rm d})= 0,
\label{density}
\end{equation}

\begin{equation}
\frac{\partial m_{\rm p}}{\partial t}+ v_{\rm r}\frac{\partial m_{\rm p}}{\partial r}= \frac{2\sqrt{\pi}a^2\Delta v_{\rm pp}}{h_{\rm d}}\Sigma_{\rm d},
\label{size}
\end{equation}
where $a=(3m_{\rm p}/4\pi\rho_{\rm int})^{1/3}$ is the particle radius, $v_{\rm r}$ is the radial drift velocity of the particles, $\Delta v_{\rm pp}$ is the relative velocity of the particles, $\rho_{\rm int}$ is the internal density of dust grains, and $h_{\rm d}$ is the dust scale height.
The internal density of dust grains is fixed to be a typical value of $\rho_{\rm int}=1.4$ g cm$^{-3}$ for simplicity, while low-density dust grains, such as fluffy dust grains, decrease the internal density to $\rho_{\rm int}\sim10^{-5}-10^{-3}$ g cm$^{-3}$ \citep{oku12}.
The other $v_{\rm r}$, $\Delta v_{\rm pp}$, and $h_{\rm d}$ are described as follows.

The radial drift velocity of particles is given by \citep{ada76,wei77}
\begin{equation}
v_{\rm r} = -\frac{2{\rm St}}{1 + {\rm St}^2}{\eta}v_{\rm K}, 
\label{eq:vr}
\end{equation}
where 
\begin{equation}
    {\rm St}=\frac{\pi}{2}\frac{\rho_{\rm int}a}{\Sigma_{\rm g}}
\label{eq:stokes}
\end{equation}
is the Stokes number, and
\begin{equation}
\eta = - \frac{1}{2}\left(\frac{c_{\rm s}}{v_{\rm K}}\right)^2\frac{d\ln{(c_{\rm s}^2{\rho}_{\rm g})}}{d\ln{r}}
\label{eq:eta}
\end{equation}
is a dimensionless quantity characterizing the pressure gradient of the disk gas, $c_{\rm s}$ is the sound speed,
and $v_{\rm K} = r\Omega_{\rm K}$ is the Kepler velocity, where $\Omega_{\rm K} = \sqrt{GM_\star/r^3} = 2.0\times 10^{-7} (r/1~\AU)^{-3/2}(M_\star/M_\odot)^{1/2}~{\rm s^{-1}}$
is the Keplerian frequency with $G$, $M_\star$ being the gravitational constant and central stellar mass, respectively. 
In our disk model, ${\eta}v_{\rm K} = 33~{\rm m~s^{-1}}$ is derived.

The particle collision velocity $\Delta v_{\rm pp}$ is given by
\begin{equation}
\Delta v_{\rm pp}=\sqrt{(\Delta v_{\rm B})^2+(\Delta v_{\rm r})^2+(\Delta v_{\phi})^2+(\Delta v_{\rm z})^2+(\Delta v_{\rm t})^2},
\label{collision}
\end{equation}
where $\Delta v_{\rm B}, \Delta v_{\rm r},\Delta v_{\phi},\Delta v_{\rm z}$, and $\Delta v_{\rm t}$ are the relative velocities induced by Brownian motion, radial drift, azimuthal drift, vertical settling, and turbulence, respectively. 
The detail equations of each component are shown in Appendix \ref{turbulence}.

The dust scale height is determined by a balance between vertical settling and turbulent diffusion and is written as
\citep[e.g.,][]{dub95,sch04,you07}
\begin{equation}
        h_{\rm d}=\Big(1+\frac{\rm St}{\alpha_{\rm D}}\frac{1+2 \rm St}{1+\rm St}\Big)^{-1/2} h_{\rm g},
\label{eq:scaleheight}
\end{equation}
where $h_{\rm g}=c_s/\Omega_{\rm K}$ is the scale height of the gas. 
The turbulence parameter $\alpha_{\rm D}$ is set to be a typical value of $\alpha_{\rm D}=10^{-3}$ in this study.

By taking into account the radial drift (Equation \ref{eq:vr}) and particle collision velocity (Equation \ref{collision}), we calculate the evolution of the dust coagulation model.

\subsection{Dust Evolution}\label{evolution}

In the previous subsection, we describe the dust coagulation model. 
Here, we show the evolution of the surface density and particle size.
We focus mainly on the dust grains that are sensitive to millimeter-wave emission.

\begin{figure}[htbp]
\includegraphics[width=8.cm,bb=0 0 1271 1618]{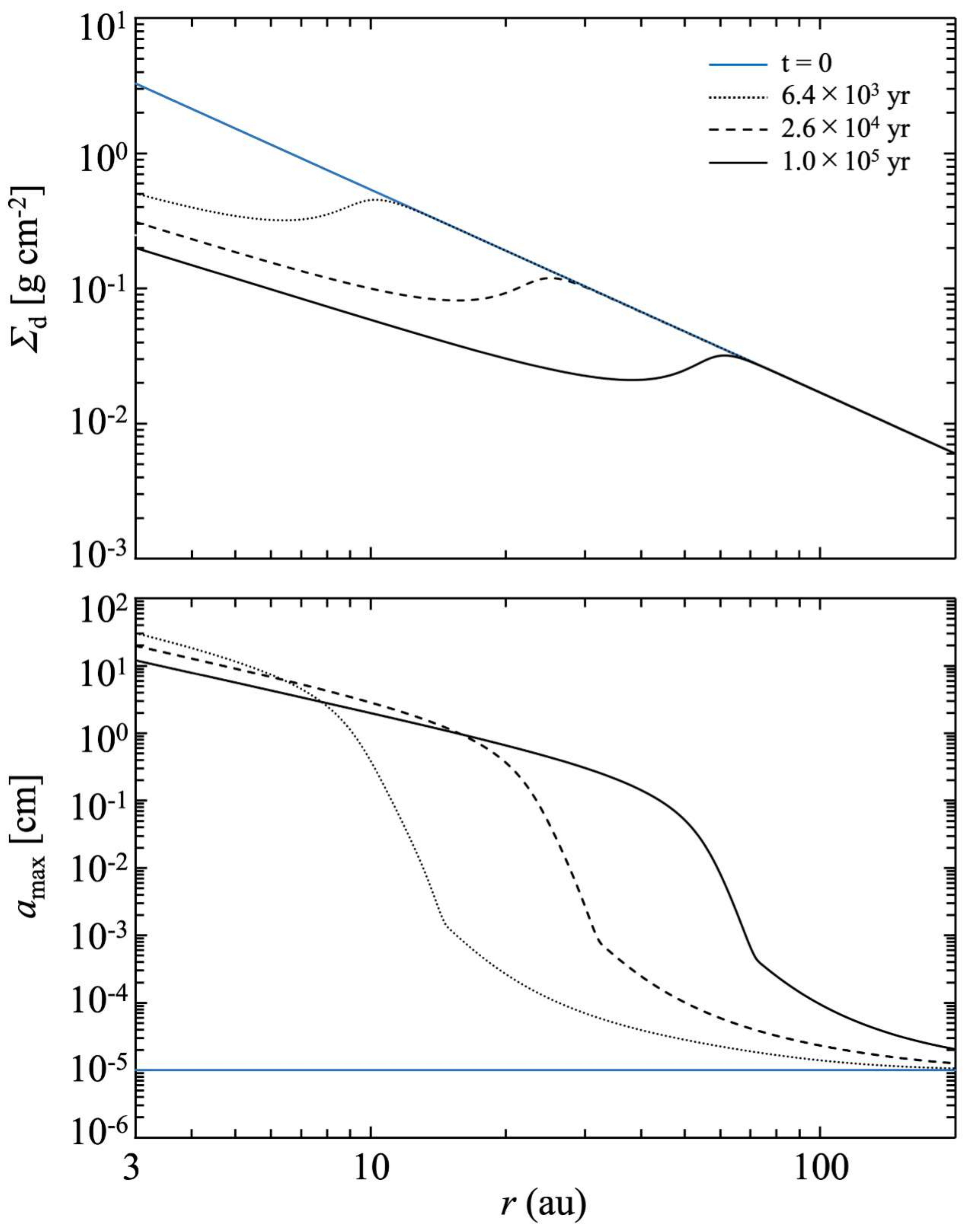}
\caption{Time evolution of the surface density $\Sigma_d$ ({\it top panel}) and radius $a$ ({\it bottom panel}) of dust particles as a function of orbital radius $r$ for models with $a=10^{-3}$.
The blue lines show the initial conditions, while the black dotted, dashed, and solid lines are the snapshots at times $t=6.4\times10^3$, $2.6\times10^4$, and $1.0\times10^5$ yr, respectively.
}
\label{radial_plot}
\end{figure}

Figure \ref{radial_plot} shows the results of the global evolution of the dust surface density $\Sigma_{\rm d}$ and particle size $a$.
In this figure, the critical radius  where the surface density remains the initial state on the outside, can be identified. Furthermore, this position moves outwards over time.
We regard this position as a growth front, where the dust evolution proceeds.
For example, the growth front corresponds to $\sim10$ au and $\sim24$ au, at $t=6.4\times10^3$ yr and $t=2.6\times10^4$ yr, respectively.
The growth front is also known as the pebble production line as suggested by \citet{lam14} because the disk particle have just grown to pebble sizes in this position.
Importantly, we point out that the position of the growth front (the pebble production line) is independent of the dust structures, disk mass, temperature, or the turbulence strength as explained in the following discussion.

As demonstrated by many previous studies \citep[e.g.,][]{tak05,gar07,bra08,bir10,bir12,oku12}, the dust evolution can be estimated from growth timescale.
The  growth rate of the aggregate mass $m$ at the midplane
is given \citep{tan05} by
\beq
\frac{dm}{dt}= \rho_{\rm d}\sigma_{\rm col}\Delta v =  \frac{\Sigma_{\rm d}\sigma_{\rm col}\Delta v}{\sqrt{2\pi} h_{\rm d} },
\label{eq:mdot}
\eeq
where $\rho_{\rm d}=\Sigma_{\rm d}/(\sqrt{2\pi}h_{\rm d})$ is the spatial dust density at the midplane, $\sigma_{\rm col}$ is the collisional cross section of two dust particles.
Then, Equation (\ref{eq:mdot}) can be rewritten in terms of the growth
timescale as
\beq
t_{\rm grow}=\left(\frac{m}{\dot{m}}\right)=  \frac{ m\sqrt{2\pi} h_{\rm d} } {\Sigma_{\rm d}\sigma_{\rm col}\Delta v}=\frac{4\sqrt{2\pi}}{3} \frac{h_{\rm d}}{\Delta v}\frac{ \rho_{\rm int} a}{\Sigma_{\rm d}},
\label{eq:tgrow_1}
\eeq
where $m=(4\pi/3) \rho_{\rm int}a^3$ and $\sigma_{\rm col}=\pi a^2$.

Here, we focus on the millimeter sized dust grains because these grains are sensitive to millimeter-wave emission.
For millimeter-sized dust grains, $h_{\rm d} \sim \sqrt{\alpha_{\rm D}/ {\rm St}} h_{\rm g}$ and 
$\Delta v \sim \Delta v_{\rm t} \sim \sqrt{\alpha_{\rm D} {\rm St}} c_{\rm s}$ \citep{bra08}. Then, we obtain 
\beq
{t_{\rm grow}} \sim \left(\frac{\Sigma_{\rm g}}{\Sigma_{\rm d}}\right)\frac{1}{\Omega_{\rm K}}.
\label{eq:tgrow}
\eeq

Equation (\ref{eq:tgrow}) indicates that the dust evolution commences from inside out because the growth timescale is roughly proportional to the orbital period. Furthermore, the growth time scale does not depend on the other parameters such as the internal density of dust grains ($\rho_{\rm int}$) and turbulence ($\alpha_{\rm D}$) excepting for the dust to gas mass ratio ($\Sigma_{\rm g}/\Sigma_{\rm d}$).
Therefore, the dust growth time is independent of the fluffiness of dust, the disk mass, temperature, or the strength of the disk turbulence, while it will become shorter by increasing the dust mass ratio.
Outside of the growth front, the dust particles still remain the initial state since they are not evolved yet, while the dust grains are grown inside of the growth front with drifting radially.
As a result, the dust surface density is maximized in the growth front (see Figure \ref{radial_plot}).
Therefore, a ring structure is expected to be observed at the growth front.
However, it should be noted that the ring-structure of the growth front has not been shown so far even though the existence of the growth front is well known.

\subsection{Dependence on the Initial Conditions}

We investigate the dependence of the growth front on the initial conditions.
Even though the MMSN model is applied for the disk evolution in the previous subsection, the protostellar disks in Class 0/I stage may have higher accretion rate and higher surface density.
Therefore, we calculate the disk evolution by changing the radial drift velocity and surface densities of gas and dust in the initial conditions.

\begin{figure}[htbp]
\includegraphics[width=8.cm,bb=0 0 1271 1618]{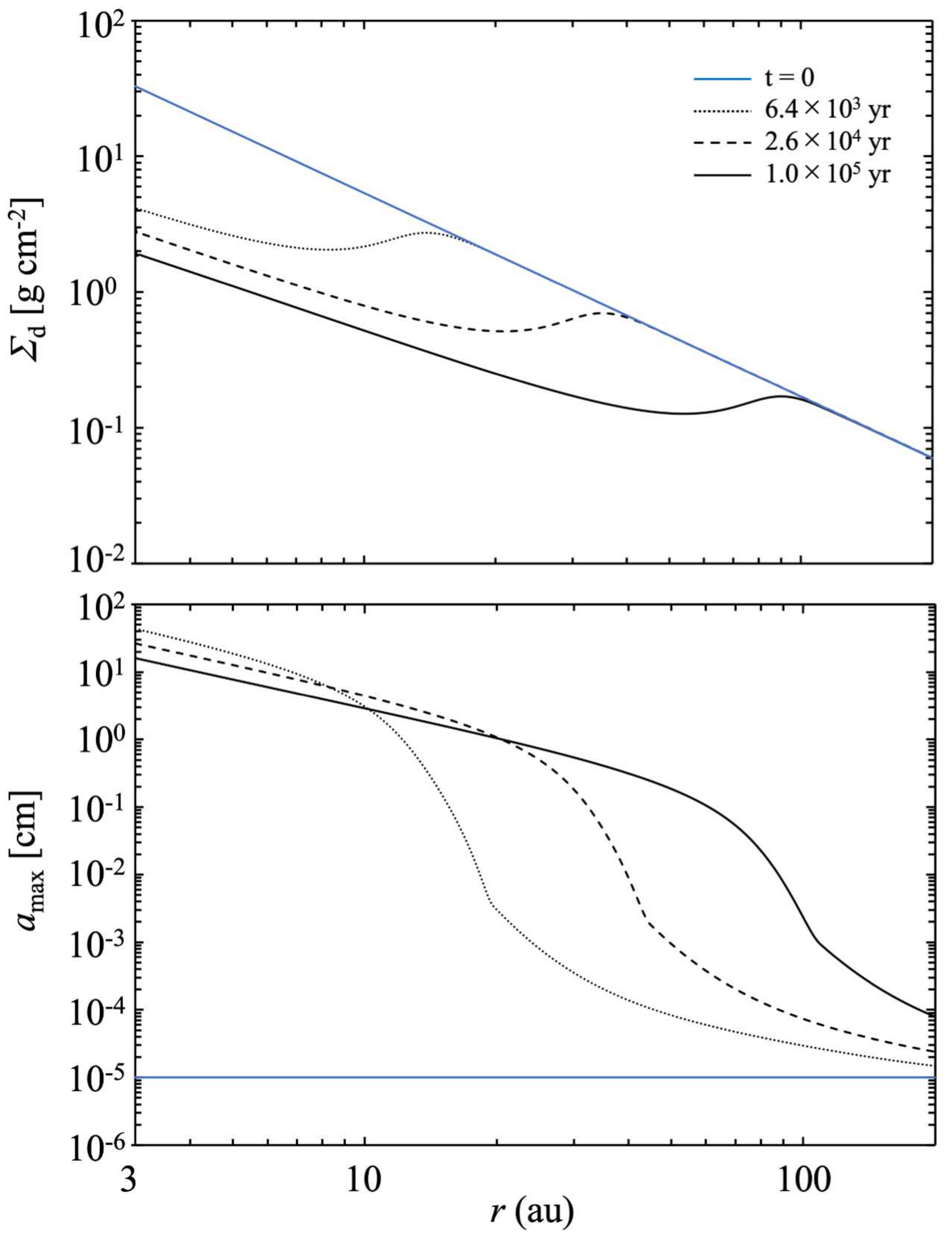}
\caption{Same as Figure \ref{radial_plot}, but the gas and dust surface densities are 10 times larger than the MMSN model and the radial drift velocity ($v_{\rm r}$) is also 10 times higher than the MMSN model as ${\eta}v_{\rm K}$=330 m s$^{-1}$.
}
\label{10x}
\end{figure}

\begin{figure}[htbp]
\includegraphics[width=8.cm,bb=0 0 1271 1618]{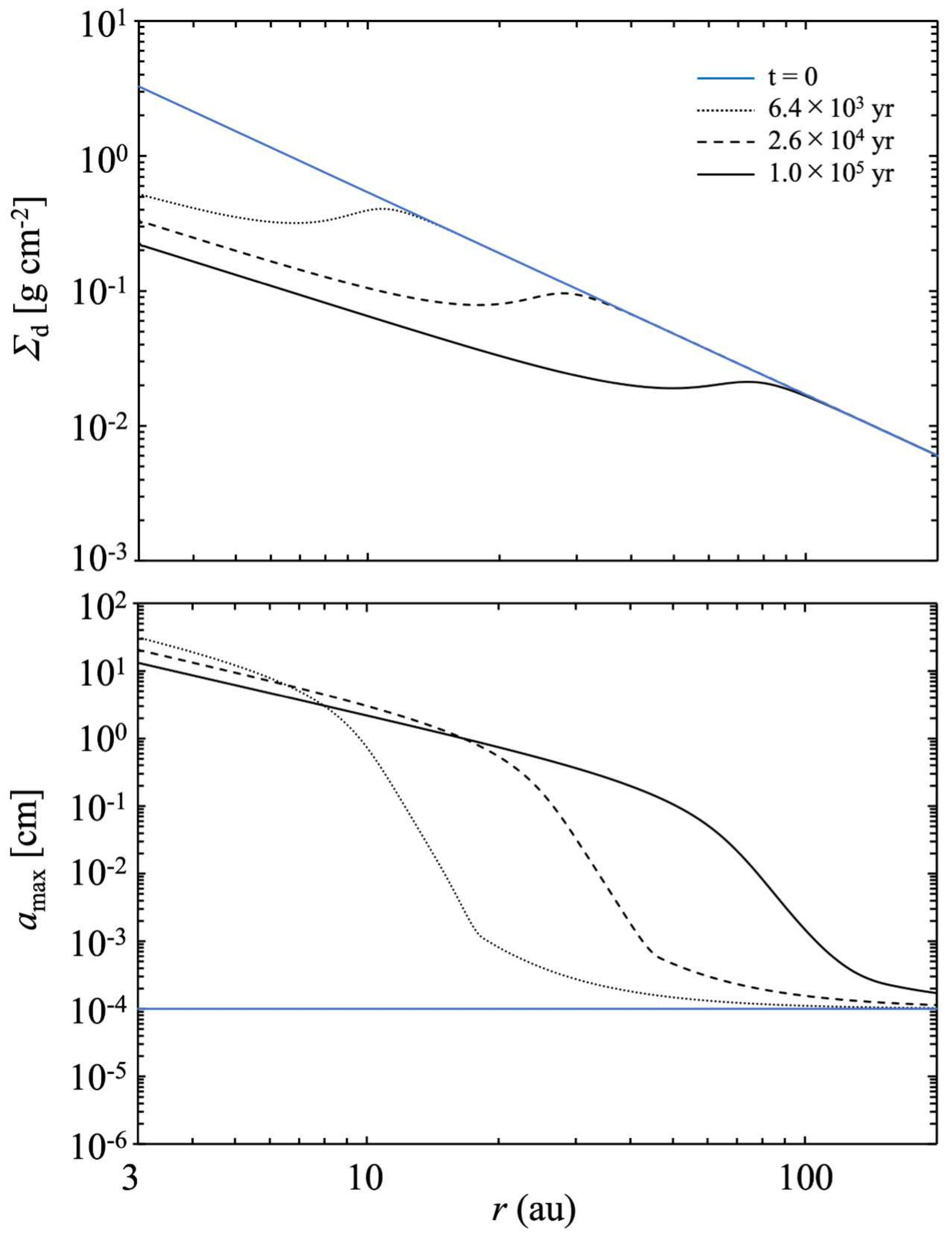}
\caption{Same as Figure \ref{radial_plot}, but the initial size of the dust particles is  1 $\mu$m.
}
\label{10amax}
\end{figure}

Figure \ref{10x} shows the results of  the case that the gas and dust surface densities are 10 times higher than the MMSN model. The dust to gas mass ratio keeps to be 1\%. 
Furthermore, to investigate the dependence of the radial drift velocity ($v_{\rm r}$) on the growth front, we use 10 times higher the value of ${\eta}$ given in equation (\ref{eq:eta}). Thus, ${\eta}v_{\rm K}$=330 m s$^{-1}$ is used.
Figure \ref{10amax} shows the disk evolution with the initial grain size of 1 $\mu$m, which is the case that the initial grain size is 10 times larger than the previous MMSN model.
By comparing the Figures of \ref{radial_plot}, \ref{10x}, and \ref{10amax}, we find that the growth front and radial profiles of dust, gas, and grain size are hardly changed with changing the initial conditions.

The growth front is appeared independently on the initial surface densities, radial velocity, and grain size.
This is consistent with the result that the growth time scale depends only on the dust to gas mass ratio ($\Sigma_{\rm g}/\Sigma_{\rm d}$) and the orbital period ($\Omega_{\rm K}$) shown in Equation (\ref{eq:tgrow}).
Therefore, the disk evolution and growth front would be robust once the Kepler disk is formed even if  surrounding materials accrete onto the disk from the envelope. 

\subsection{Dust Ring Structure at the Growth Front by Radiative Transfer Calculations}
In this subsection, we demonstrate that a disk with a growth front can be observed as a ring by using radiative transfer calculations with RADMC-3D\footnote{RADMC-3D is an open code of radiative transfer calculations developed by Cornelis Dullemond. The code is available online at: \url{http://www.ita.uniheidelberg.de/~dullemond/software/radmc-3d/}} \citep{dul12}. 
The physical structure of the disk are shown in the previous subsection and Figure \ref{radial_plot}.
The calculation setup for the radiative transfer is described as follows.

We analyze the disk with two different observing wavelengths to investigate the dependence of the growth front on the observing wavelength.
The observing wavelengths are set to $\lambda=870$ $\mu$m and 7 mm corresponding to ALMA Band 7 and VLA Q band observations, respectively.
The distance is assumed to be 100 pc.
The intensity is calculated by the radiative transfer equation using the given dust surface density, temperature, and dust absorption/scattering opacity.

The dust opacity is calculated using Mie theory. 
We calculate the dust evolution under the assumption of a power-law size distribution with an exponent of $q = -3.5$.
If collisional velocities are independent of the masses of colliding dust particles, collisional cascade via erosive collisions results in a power-law size distribution with an exponent of $q = -3.5$ \citep{doh69,tan96}. 
Although the index $q$ is modified due to the mass dependence of velocity \citep{kob10}, we put $q = -3.5$ for simplicity.
Note that the collisional sticking effectively occurs for $\Delta v \la 80$ m/s \citep{wad13}, and this condition is satisfied in the entire disk for our calculations.

\begin{figure}[htbp]
\includegraphics[width=8.cm,bb=0 0 2226 1521]{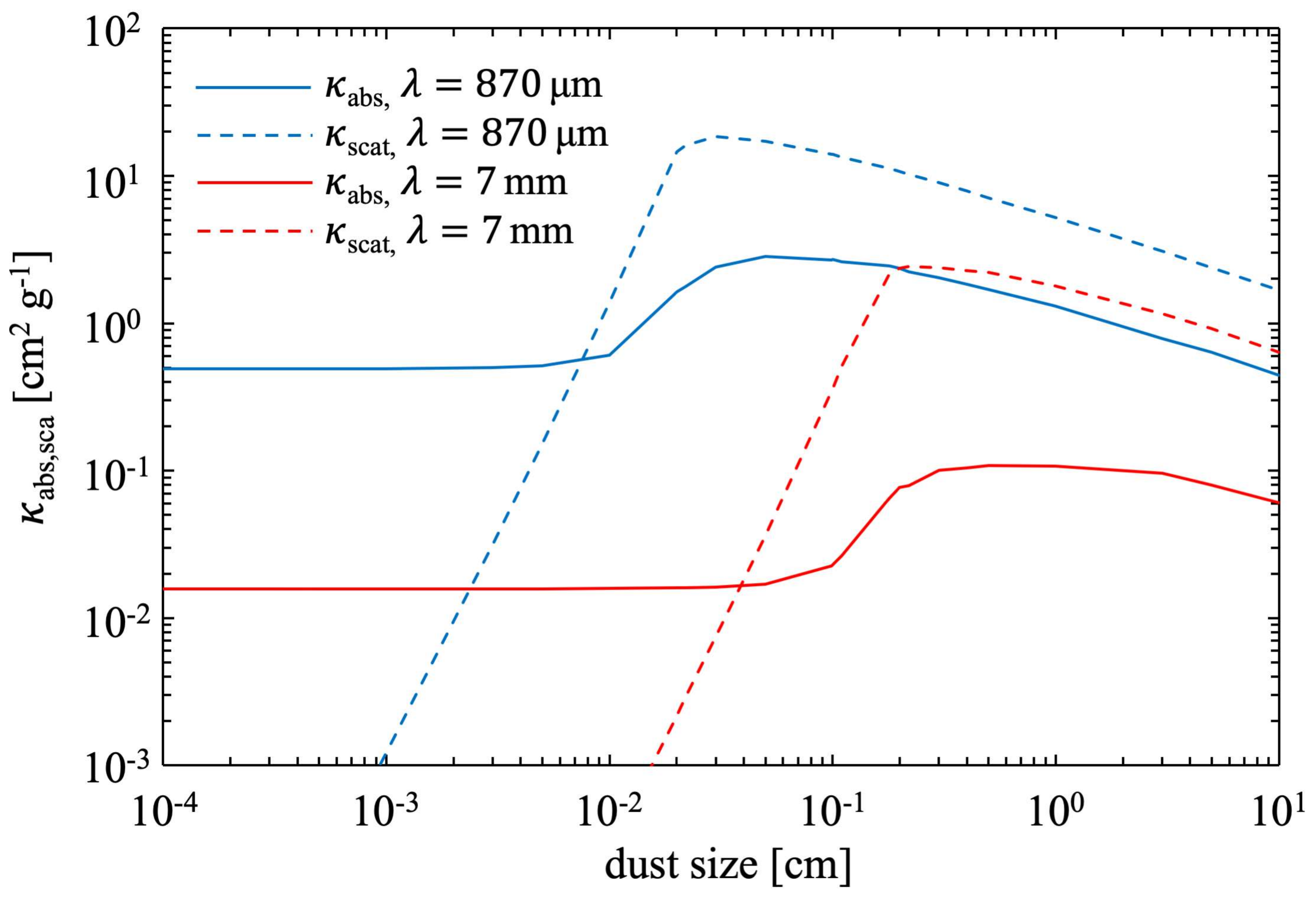}
\caption{Absorption and scattering mass opacity of dust grains in the cases
of $\lambda=870$ $\mu$m and 7 mm. The size distribution is a power-law with an exponent of $q=-3.5$.}
\label{opacity}
\end{figure}

The dust composition was assumed to be a mixture of silicate (50\%) and water ice (50\%) \citep{pol94}. We used the refractive index of astronomical silicate \citep{wei01} and water ice \citep{war84} and calculated the absorption and scattering opacity based on effective medium theory using the Maxwell-Garnett rule \cite[e.g.,][]{boh83,miy93}. 
Figure \ref{opacity} shows absorption and scattering opacity as a function of a dust grain size.

\begin{figure*}[htbp]
\includegraphics[width=18.cm,bb=0 0 2865 1638]{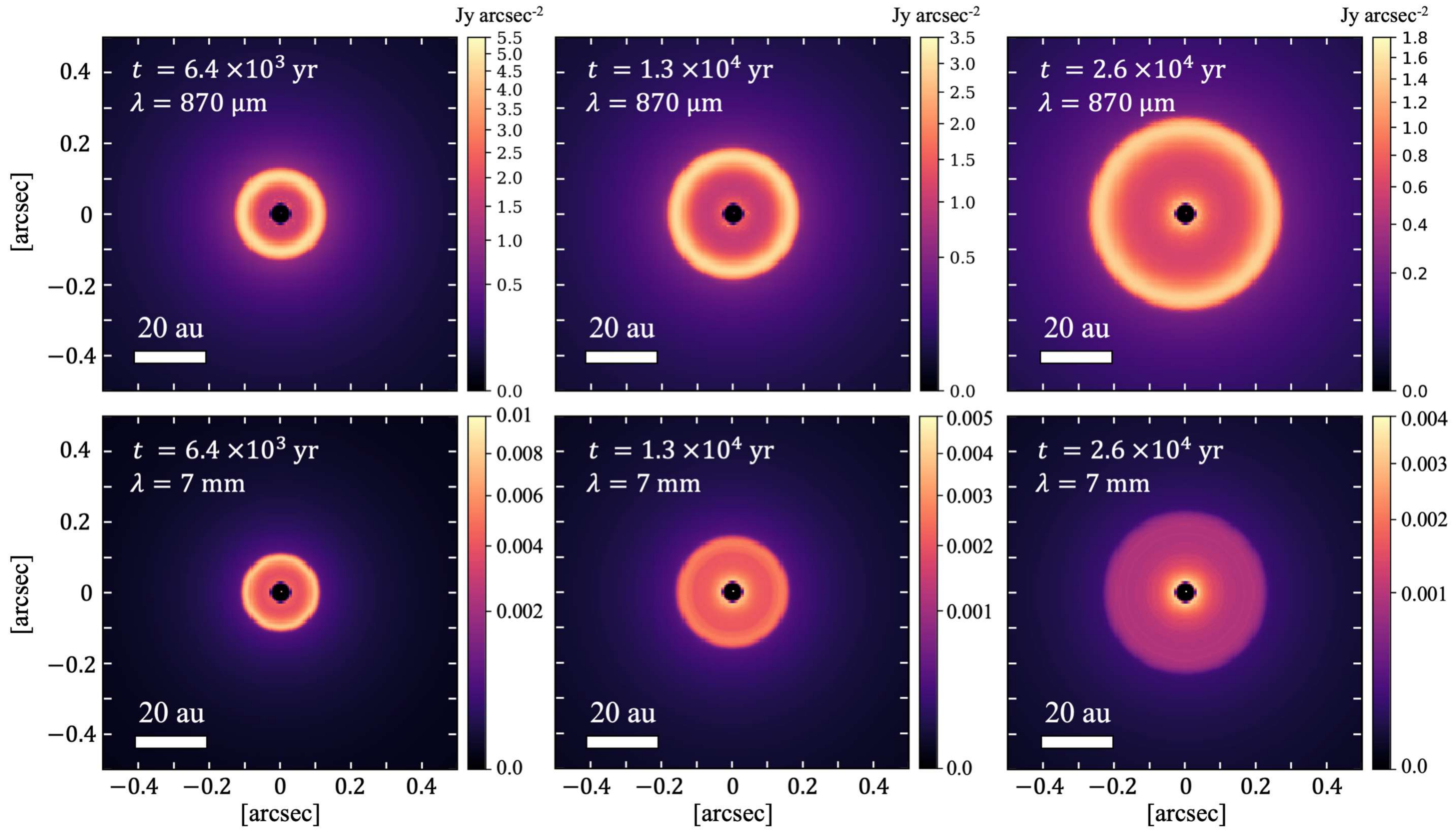}
\caption{The intensity maps of the dust coagulation model at $t=6.4\times10^3$ yr, $1.3\times10^4$ yr, and $2.6\times10^4$ yr obtained by the radiative transfer calculations with RADMC-3D. The observing wavelengths are $\lambda=870$ $\mu$m (upper panel) and 7 mm (lowe panel), respectively.
The distance is 100 pc.}
\label{image}
\end{figure*}

The radiative transfer calculation is performed without changing the dust scale height for each dust size. If the disk is face-on, the variations of the dust scale height will be less effected.
Here, we assume that the dust scale height is an order of magnitude thinner than the gas scale height.

Figure \ref{image} shows images of the radiative transfer calculations of our model at $t=6.4\times10^3$ yr, $t=1.3\times10^4$ yr, and $t=2.6\times10^4$ yr.
We find that the ring structure can be observed at the growth front with both wavelengths. The ring position moves outward over time.

By comparing the 870 $\mu$m and 7 mm images, we find that the ring structure is observed at a few au inner radius in the 7 mm image more than in the 870 $\mu$m image.
We also find that the ring structure is more enhanced in the 870 $\mu$m image than the 7 mm image.
These different images are caused by the different observing wavelengths. 
The growth front has the millimeter-sized grains.
The observing wavelength at 870 $\mu$m is sensitive to millimeter-sized grains, whereas the 7 mm wavelength is sensitive to centimeter-sized grains. 
Therefore, the 870 $\mu$m image indicate the ring structure better than the 7 mm image.
The different ring positions between the 870 $\mu$m and 7 mm images indicate that the observed ring is caused by the combined effect of the local maximum of the surface density and the millimeter-sized grains at the growth front.

\subsection{Spectral Index Distributions}
In the previous subsection, we show that the size of dust grains changes across the growth front.
One way of measuring the dust grain size is to derive the frequency dependence of thermal dust continuum emission since larger grains efficiently emit thermal radiation at a wavelength similar to their size \citep[e.g.,][]{dra06}.
Therefore, the spectral index, $\alpha$, provides us information on grain sizes.

We investigate the spectral index across the growth front by using the model images shown in Figure \ref{image}.
The spectral index $\alpha$ is calculated as
\begin{equation}
\alpha=\frac{\ln I_1 - \ln I_2}{\ln \nu_1 - \ln \nu_2},
\label{eq:alpha}
\end{equation}
where $I$ and $\nu$ are intensity and frequency at each band. Here, we use the intensity maps of the 870 $\mu$m and 7 mm wavelengths.

\begin{figure*}[htbp]
\includegraphics[width=18.cm,bb=0 0 2817 913]{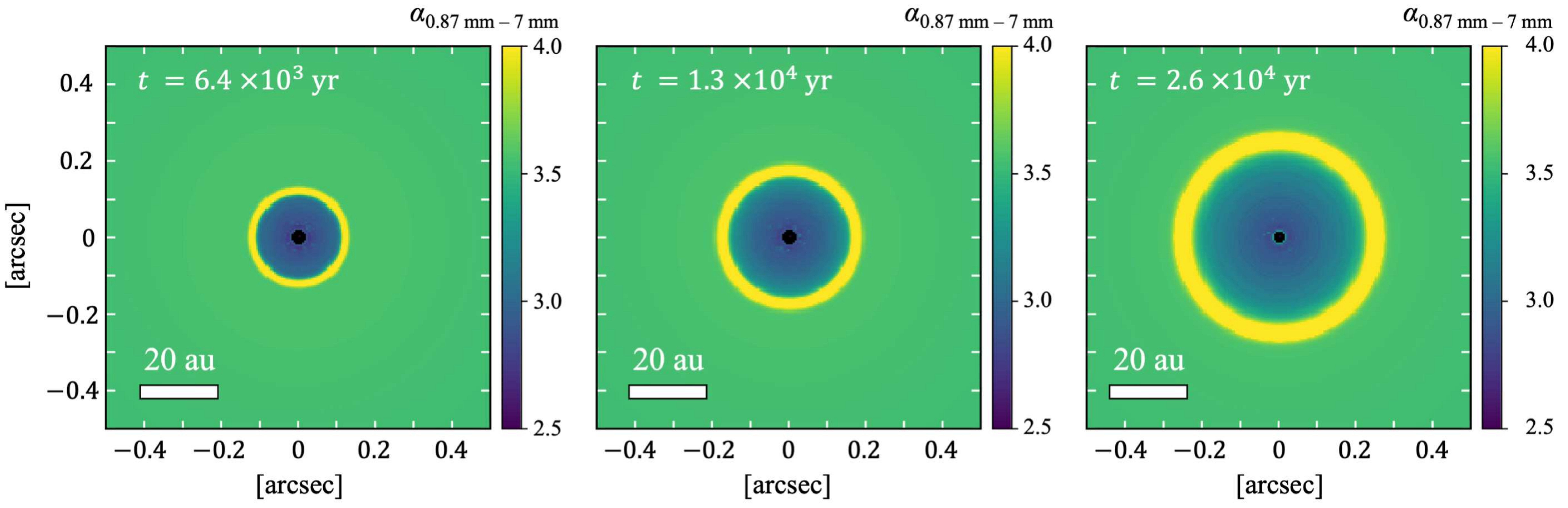}
\caption{The spectral index ($\alpha$) maps of the dust coagulation model at $t=6.4\times10^3$ yr, $1.3\times10^4$ yr, and $2.6\times10^4$ yr. The spectral index $\alpha$ is derived by the intensities of the 870 $\mu$m and 7 mm images.}
\label{image_alpha}
\end{figure*}

Figure \ref{image_alpha} shows the spectral index maps at $t=6.4\times10^3$ yr, $1.3\times10^4$ yr, and $2.6\times10^4$ yr.
These images have the same trend among time evolution.
Inside the growth front, the spectral index shows $\alpha_{\rm 0.87 mm - 7 mm}\sim2.9$, while it shows $\alpha_{\rm 0.87 mm - 7 mm}\sim3.6$ outside the growth front.
Furthermore, the spectral index takes the peak value of $\alpha_{\rm 0.87 mm - 7 mm}\sim4.1-4.4$ at the growth front because the growth front has the millimeter-sized dust grains that efficiently emit the thermal radiation at the 860 $\mu$m observing wavelengths.
According to Figure 3 of \citet{ric10}, dust grains with a size of $< 100$ $\mu$m show a spectral index of $\beta\sim1.7$, those with a size of $\sim0.1-1$ mm show $\beta\sim2-3$, and those with a size of $> 1$ mm show $\beta<1.5$. 
Note that the observed (sub)millimeter spectral index ($\alpha$) is related to $\beta$ by $\alpha=\beta+2$ in the Rayleigh–Jeans limit.
Thus, these $\alpha$ values are consistent with the grain sizes of our dust model.
Therefore, the spectral index is the possible way to identify the growth front.

It should be noted that the estimates of the grain sizes from the spectral index depend on the dust model such as dust chemical composition, power law of dust size distribution, porosity and so on. 
Even though the absolute value of the spectral index, $\alpha$, changes by dust model, the behavior of the spectral index is robust.

\subsection{Growth Front Location}

\begin{figure}[htbp]
\includegraphics[width=8.cm,bb=0 0 1677 1173]{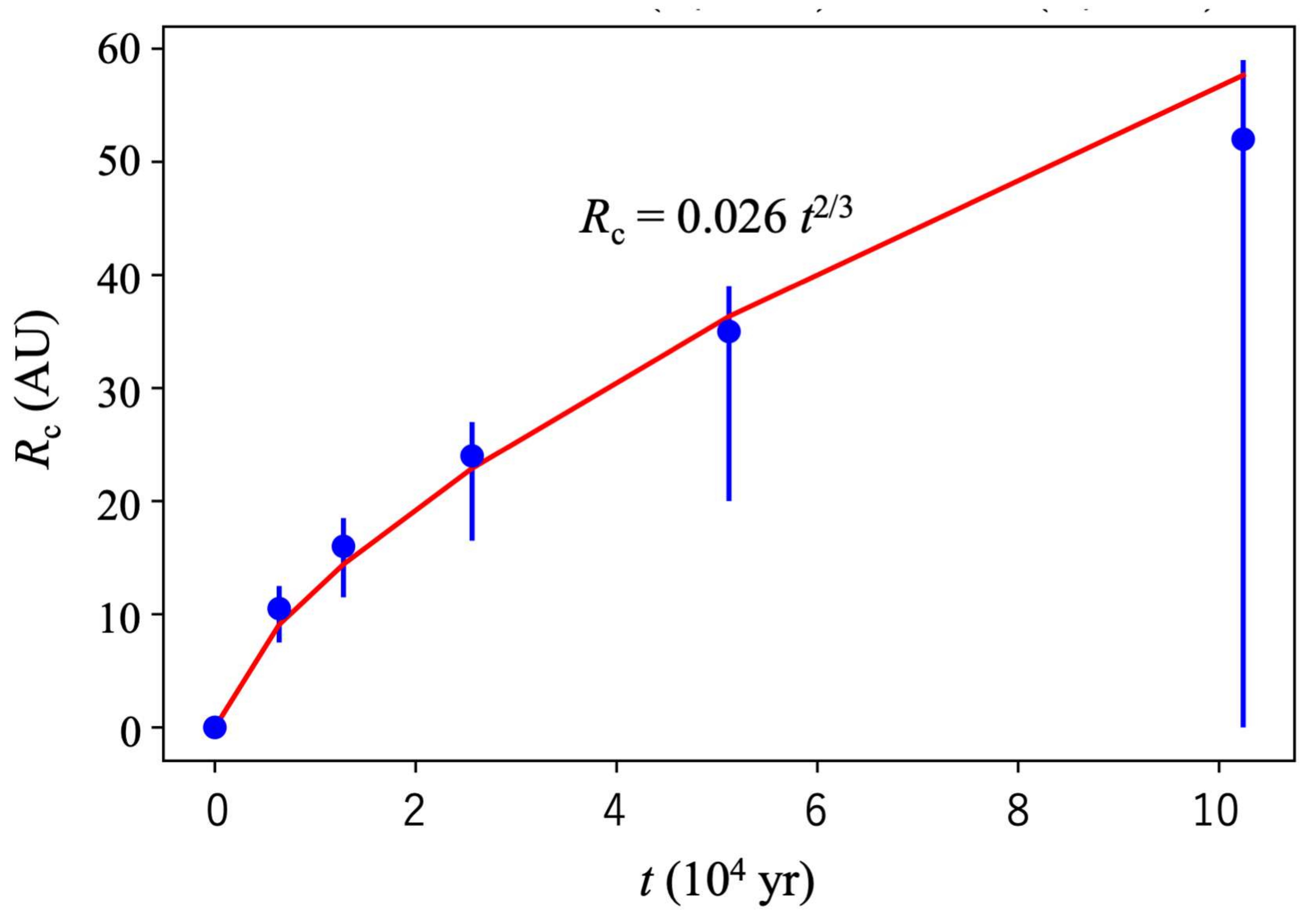}
\caption{The ring positions of our model against time. The red line indicates the fitting to equation (\ref{eq:front}) to derive the transformation coefficient $A$.
}
\label{Acoeff}
\end{figure}

We formulate  the ring location.
Equation (\ref{eq:tgrow}) indicates the timescale of the dust evolution which is a function of the Keplerian frequency $\Omega_{\rm K}$.
In other words, if the timescale ($t_{\rm grow}$) is set to the disk age ($t_{\rm age}$), we can derive the critical radius ($R_{\rm c}$) where the growth front reaches because $\Omega_{\rm K}\propto r^{-3/2}$.
The critical radius can be estimated by transforming the equation (\ref{eq:tgrow}) into a function of $r$.
Thus, Equation (\ref{eq:tgrow}) yields
\beq
R_{\rm c}=A\left(\frac{M_\star}{M_\odot}\right)^{1/3}\left(\frac{\zeta_{\rm d}}{0.01}\right)^{2/3}\left(\frac{t_{\rm disk}}{1\ {\rm yr}}\right)^{2/3}\ {\rm au}
\label{eq:front}
\eeq
where $A$ is the transformation coefficient and $\zeta_{\rm d}=\Sigma_{\rm d}/\Sigma_{\rm g}$.
The coefficient $A$ can be derived by fitting the various ring positions in time to Equation (\ref{eq:front}).
Therefore, we perform the radiative transfer calculations for the model at $t=6.4\times10^3$ yr, $1.3\times10^4$ yr, $2.6\times10^4$ yr, $5.2\times10^4$ yr, and $1.0\times10^5$ yr by assuming $\lambda=870$ $\mu$m, $M_\star=1 M_{\rm \odot}$, and $\zeta_{\rm d}=0.01$. Then, we identify the ring position.
Figure \ref{Acoeff} shows the time evolution of the ring position.
The error of the ring position indicates the full width half maximum (FWHM)  derived by the direct measurements of the synthetic images.
By fitting the ring positions to equation (\ref{eq:front}), $A$ is derived to be $0.026$.

\section{Observational Study of the Growth Front} \label{sec:discussion}

In the previous section, we show that the dust coagulation model is one possible scenario for the formation of the dust ring.
The dust ring position (growth front) moves outward over time because the growth timescale is roughly proportional to the orbital period.
Here, we discuss the growth front and observed dust ring positions.

\subsection{Comparison with Observations}

In this subsection, we compare the growth front position with observed ring positions. Many dust ring structures have been identified with ALMA high spatial resolution observations even though the origin of such rings is still under debate.

\begin{table*}[!ht]
\centering
\vspace{-2mm}
\caption{Data sample.}
\footnotesize
\begin{tabular}{lccccc}
%\begin{tabular}{lllllllll}
\hline
\hline
Name&stellar mass ($M_\star$) & age ($t_{\rm age}$) & growth front ($R_{\rm c}$)&class & observed ring position ($R_{\rm p}$)  \\
 & ($M_\odot$) & (Myr) &  (AU)&  &   (AU) \\
\hline
L1527	&	0.5$^\alpha$	&	0.037$^\alpha$	&	22	&	0/I	&	15$\rm ^a$ 		\\
\hline
WL~17	&	0.3	&	0.1	&	38	&	0/I	&	17$\rm ^b$		\\
\hline
IRS~63	&	0.8	&	0.13	&	62	&	0/I	&	27$\rm ^c$		\\
 & & & & & 51$\rm ^c$ \\
\hline
GY~91	&	0.3	&	0.5	&	100	&	0/I	&	25$\rm ^d$ 	\\
 & & & & & 55$\rm ^d$ \\
 & & & & & 82$\rm ^d$ \\
 \hline
Elias~24	&	0.8	&	0.2	&	81	&	II	&	77$\rm ^e$	 	\\
 & & & & & 123$\rm ^e$ \\
 \hline
WaOph~6	&	0.7	&	0.3	&	110	&	II	&	88$\rm ^e$	 	\\
\hline
IM Lup	&	0.9	&	0.5	&	160	&	II	&	134$^e$	 	\\
\hline
RU Lup	&	0.6	&	0.5	&	140	&	II	&	17$\rm ^e$	 	\\
 & & & & & 25$\rm ^e$ \\
 & & & & & 34$\rm ^e$ \\
 & & & & & 50$\rm ^e$ \\
 \hline
HL Tau	&	0.5	&	0.5	&	130	&	II	&	21$\rm ^f$	 	\\
 & & & & & 40$\rm ^f$ \\
 & & & & & 49$\rm ^f$ \\
 & & & & & 58$\rm ^f$ \\
 & & & & & 72$\rm ^f$ \\
 & & & & & 85$\rm ^f$ \\
 & & & & & 102$\rm ^f$ \\
\hline
WSB~52	&	0.5	&	0.6	&	150	&	II	&	25$\rm ^e$	 	\\
\hline
Elias~20	&0.5		&	0.8	&	180	&	II	&	29$\rm ^e$	 	\\
 & & & & & 36$\rm ^e$ \\
 \hline
SR~4	&	0.7	&	0.8	&	200	&	II	&	18$\rm ^e$	 	\\
\hline
Elias~27	&0.5		&	0.8	&	180	&	II	&	86$\rm ^e$	 	\\
\hline
AS~209	&	0.8	&	1	&	240	&	II	&	14$\rm ^e$	 	\\
 & & & & & 28$\rm ^e$ \\
 & & & & & 39$\rm ^e$ \\
 & & & & & 74$\rm ^e$ \\
 & & & & & 97$\rm ^e$ \\
 & & & & & 120$\rm ^e$ \\
 & & & & & 141$\rm ^e$ \\
 \hline
Sz~114	&	0.2	&	1	&	140	&	II	&	45$\rm ^e$	 	\\
\hline
DoAr~33	&		1.1&	1.6	&	360	&	II	&	17$\rm ^e$	 	\\
\hline
DoAr~25	&	1	&	2	&	410	&	II	&	86$\rm ^e$	 	\\
 & & & & & 111$\rm ^e$ \\
 & & & & & 137$\rm ^e$ \\
 \hline
GW~Lup	&	0.5	&	2	&	320	&	II	&	85$\rm ^e$	 	\\
 & & & & & 108$\rm ^e$ \\
 \hline
HD~143006	&	1.8	&	4	&	790	&	II	&	6$\rm ^e$	 	\\
 & & & & & 41$\rm ^e$ \\
 & & & & & 65$\rm ^e$ \\
 \hline
Sz~129	&	0.8	&	4	&	610	&	II	&	10$\rm ^e$	 	\\
 & & & & & 46$\rm ^e$ \\
 & & & & & 69$\rm ^e$ \\
 \hline
MY Lup	&		1.2&	10	&	1300	&	II	&	20$\rm ^e$	 	\\
 & & & & & 40$\rm ^e$ \\
 \hline
HD~163296	&	2	&	12.6	&	1800	&	II	&	14$\rm ^e$	 	\\
 & & & & & 67$\rm ^e$ \\
 & & & & & 100$\rm ^e$ \\
 & & & & & 155$\rm ^e$ \\
 \hline
HD~142666	&	1.6	&	12.6	&	1600	&	II	&	6$\rm ^e$	\\
 & & & & & 20$\rm ^e$ \\
 & & & & & 40$\rm ^e$ \\
 & & & & & 58$\rm ^e$ \\
\hline
\end{tabular}\\
 Refs. a. \citet{nak20}, b. \citet{she18}, c. \citet{seg20}, d. \citet{she17}, e. \citet{hua18}, f. \citet{alma15}\\
$^\alpha$ The mass and age are estimated by \citet{aso17} and \citet{tak16}, respectively. 
\label{tbl:sample}
\end{table*}

We use the results of the DSHARP project \citep{and18} as \citet{hua18} listed the positions of the dust rings of the 18 disks in the DSHARP sample. In addition, our sample includes HL Tau (Class II) and 4 class 0/I objects (L1527, WL 17, IRS 63, and GY 91). These objects are reported to have possible ring structures. A total of 23 disks having ring structures are investigated.

\begin{figure}[htbp]
\includegraphics[width=8.cm,bb=0 0 2027 1604]{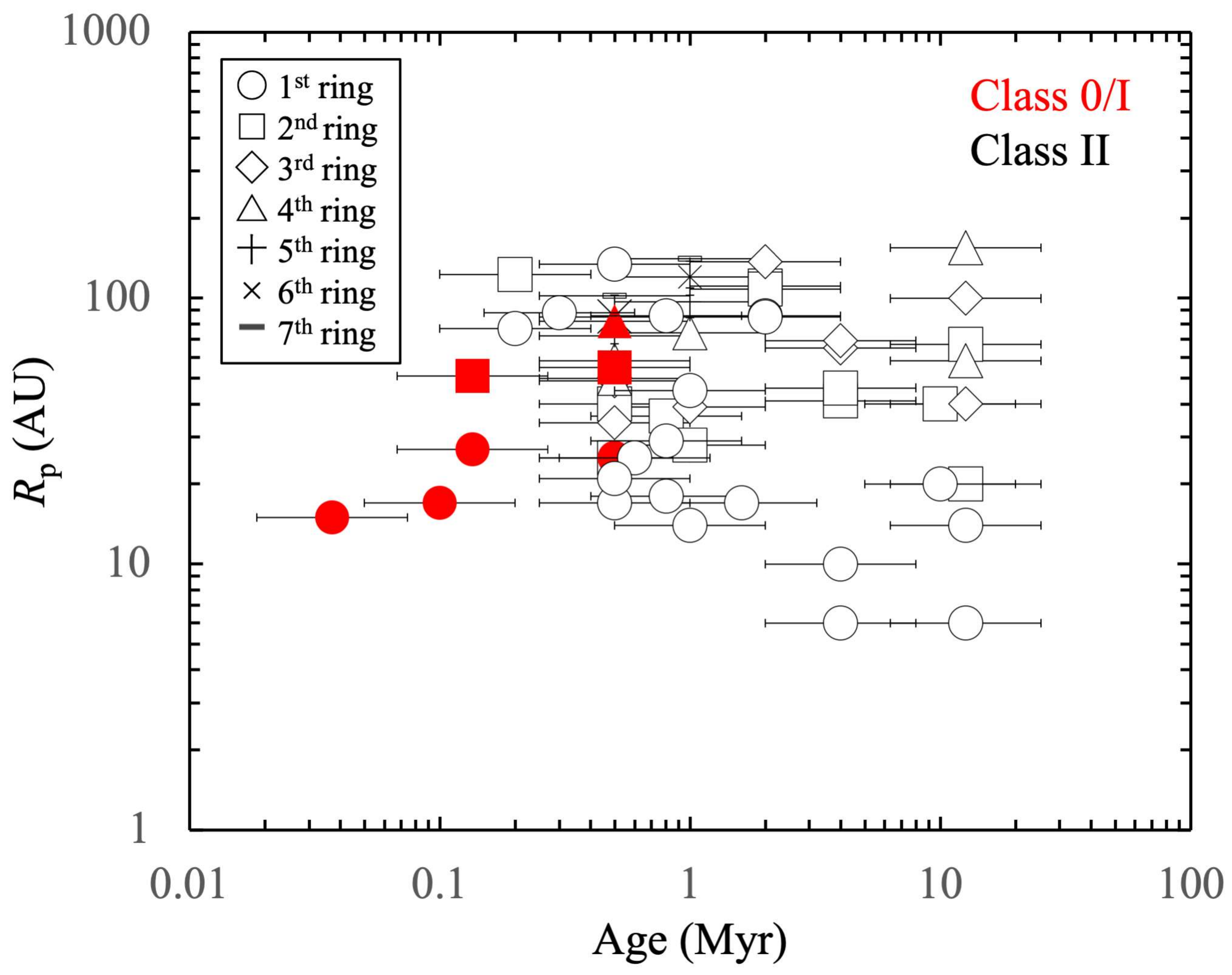}
\caption{The positions of the observed dust rings ($R_{\rm p}$) are shown against the stellar age.
Since some disks have more than one ring, the symbols in the figure are changed according to the order of the rings. Furthermore, class 0/I object rings are shown in red, and class II rings are shown as open symbols.
}
\label{fig1}
\end{figure}

Table \ref{tbl:sample} lists source name, stellar mass ($M_\star$), age ($t_{\rm age}$), and growth front radius ($R_{\rm c}$). The location of the growth front is derived by using equation (\ref{eq:front}) with assuming $t_{\rm disk}=t_{\rm age}$ and $A=0.026$.
Table \ref{tbl:sample} also lists the observed ring positions ($R_{\rm p}$). 
From these samples, we show the ring position ($R_{\rm p}$) against stellar age in Figure \ref{fig1}.
Since some disks have multiple rings, the symbols in the figure are changed according to the order of the rings.  The rings of Class 0/I objects are shown in red color and that of Class II are shown in open symbols.
The errors of the stellar age are estimated by \citet{and18} in the DSHARP sample, and we use the average value of a factor two error in all source.

As shown in Figure \ref{fig1}, we find no correlation between stellar age and ring location.
The ring location ranges from $\sim10$ to $\sim100$ au independent of stellar age.
\citet{van19} also find no evidence of snow line nor resonances of planets by investigating the gap/ring radii of 16 disks with stellar age and luminosity.
Therefore, the origin of these dust ring (and gap) structures so far remains unclear.

\begin{figure}[htbp]
\includegraphics[width=8.cm,bb=0 0 2075 1604]{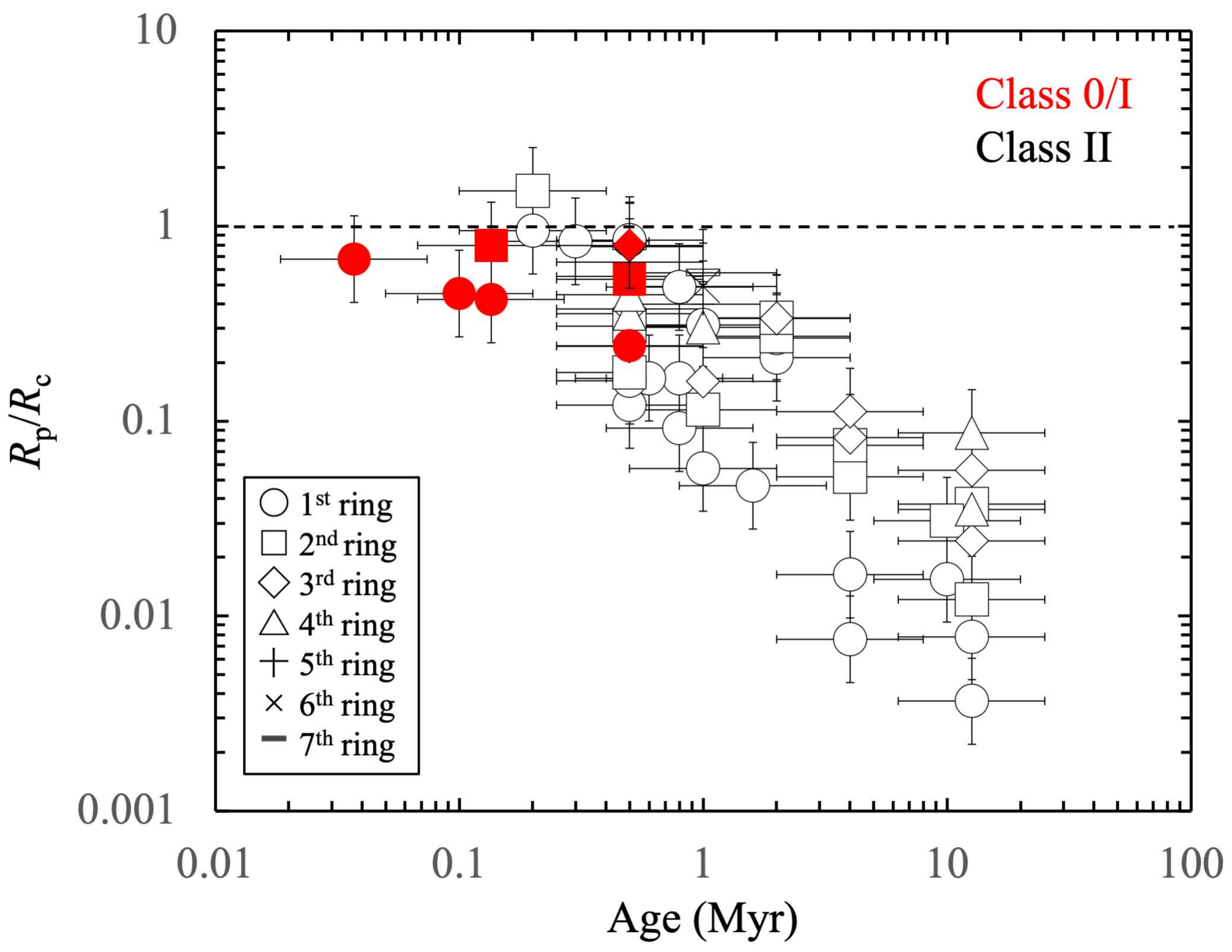}
\caption{The ratio of the ring positions ($R_{\rm p}$) to the growth front ($R_{\rm c}$) are shown against the stellar age. The ratio of unity indicates that the position of the ring corresponds to the growth front, which is found in the stellar age of $\lesssim1$ Myr. The symbols and colors represent the same with Figure \ref{fig1}.
}
\label{fig2}
\end{figure}

To investigate the origin of the ring structure by the growth front, we plot the ring location ($R_{\rm p}$) normalized by the growth front ($R_{\rm c}$) with respect to the stellar/disk age in Figure \ref{fig2}. 
This figure shows that there are some rings having the ratio ($R_{\rm p}/R_{\rm c}$) of almost unity within the disk age of less than 1 Myr, suggesting that the one of the ring positions in these young disks corresponds to the growth front.
Therefore, the growth front can explain the ring structure in particular for early stage of the disk evolution such as Class 0 and I sources.
Note that the error of $R_{\rm c}$ is calculated from the error of the stellar age because the stellar age has the largest uncertainty.

\begin{figure*}[htbp]
\includegraphics[width=18.cm,bb=0 0 2764 1018]{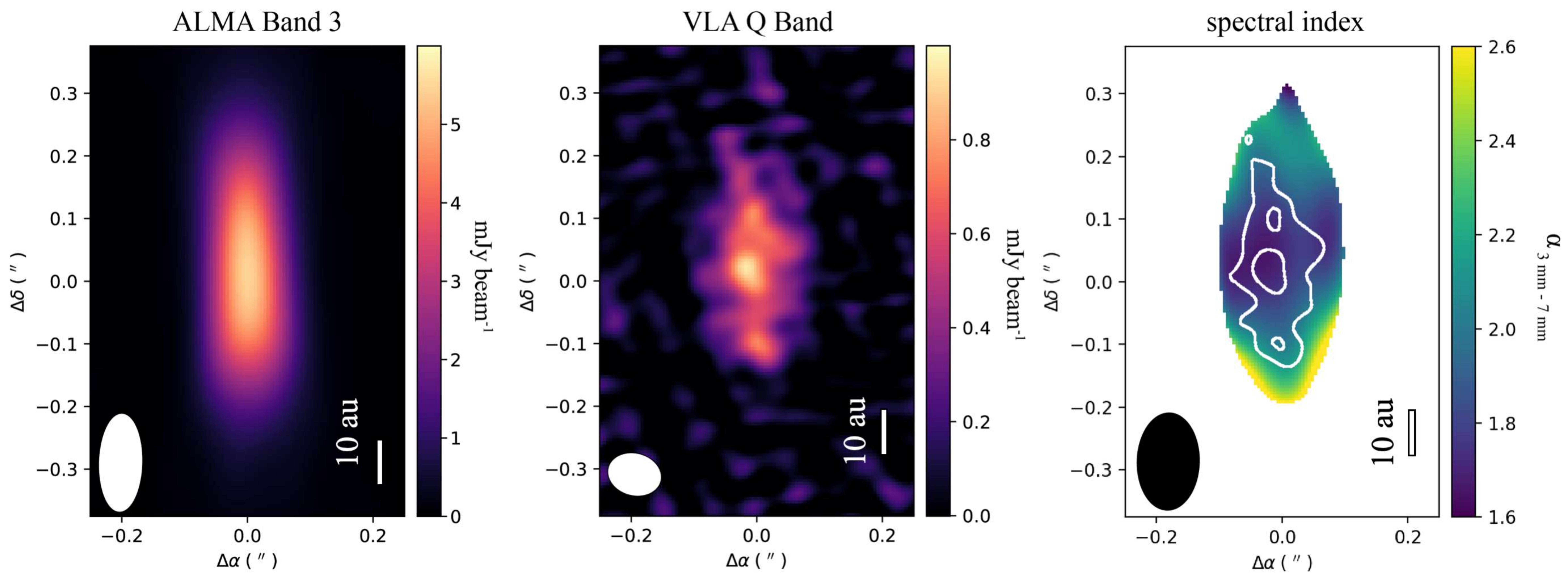}
\caption{Left and middle panels show the intensity maps of the L1527 protostar observed by ALMA Band 3 and VLA Q Band, respectively. Right panel shows the map of the spectral index $\alpha_{\rm 3 mm - 7 mm}$ derived by those ALMA Band 3 and VLA Q Band data. The white contours indicate the VLA Q Band emission distribution. The fist contour starts at 3.5$\sigma$ ($\sigma=0.11$ mJy beam$^{-1}$) and the interval is 3$\sigma$.
The white/black ellipses at the bottom left in each penal indicate the beam sizes.
The stellar position is set to 4$^h39^m53^s$.876 $+26^{\circ}3'9\farcs$47.
}
\label{l1527}
\end{figure*}

In contrast, the ratio, $R_{\rm p}/R_{\rm c}$, decreases after 1 Myr, indicating that the growth front extends much larger than the observed ring positions.
The growth front is even larger than the disk radii in several protoplanetary disks.
These observed ring structures cannot be explained by the growth front.
Thus, we suggest that the origin of the rings would be different from the growth front in the protoplanetary disks around Class II objects after 1 Myr.

\subsection{A Case Study of the Growth Front: L1527}

In the previous subsection, we find that the growth front shows the ring structure and is consistent with the observed ring positions in particular for the disks around Class 0 and I sources.
Here, we investigate the dust coagulation in more detail in an ideal disk where an observed dust ring is consistent with a growth front.
We expect that dust size would be different between inside and outside the growth front.
Inside the growth front, dust grains are expected to be larger.
In contrast, outside the growth front, dust growth is not yet proceeded and dust size is expected to be small.
Therefore, the spectral indices have different values across the growth front as shown in Figure \ref{image_alpha}.

We consider that the disk in the Class 0/I source of L1527 is an ideal target to investigate the dust coagulation because the ring position ($R_{\rm p}=15$ au) is consistent with the growth front ($R_{\rm c}=22^{+15}_{-9}$ au) within the error and also because observations over a wide range of wavelengths were performed toward this source.
L1527 is well studied by many observations with ALMA and VLA, and revealed to be forming a Keplarian disk with a radius of 80 au \citep[e.g.,][]{tob12,oha14,aso17,sak14,sak19}.
\citet{nak20} found substructures within the Keplarian disk by using VLA 7 mm dust continuum observations and interpreted this structure as a dust ring even though the disk is almost edge-on view.

We show the L1527 images of ALMA Band 3 and VLA Q Band continuum data in Figure \ref{l1527}.
The wavelengths of ALMA Band 3 and VLA Q Band are 3 mm and 7 mm, respectively.
The spectral index $\alpha_{\rm 3 mm - 7 mm}$ map is also shown in Figure \ref{l1527} and is discussed later in this section.
Detailed morphology of these continuum emission is described in \citet{nak20}.
Here, we point out that the VLA Q Band image indicates equally spaced clumps along the north-south direction at a distance of 15 au from the central protostar, which is interpreted as a ring structure in the edge-on disk.
 The ALMA Band 3 image shows no sign of the substructure, which would be due to the large beam size of the ALMA Band 3 observations.  The optical depth may also affect the continuum image if the emission is optically thick.

\begin{figure}[htbp]
\includegraphics[width=8.cm,bb=0 0 2967 1448]{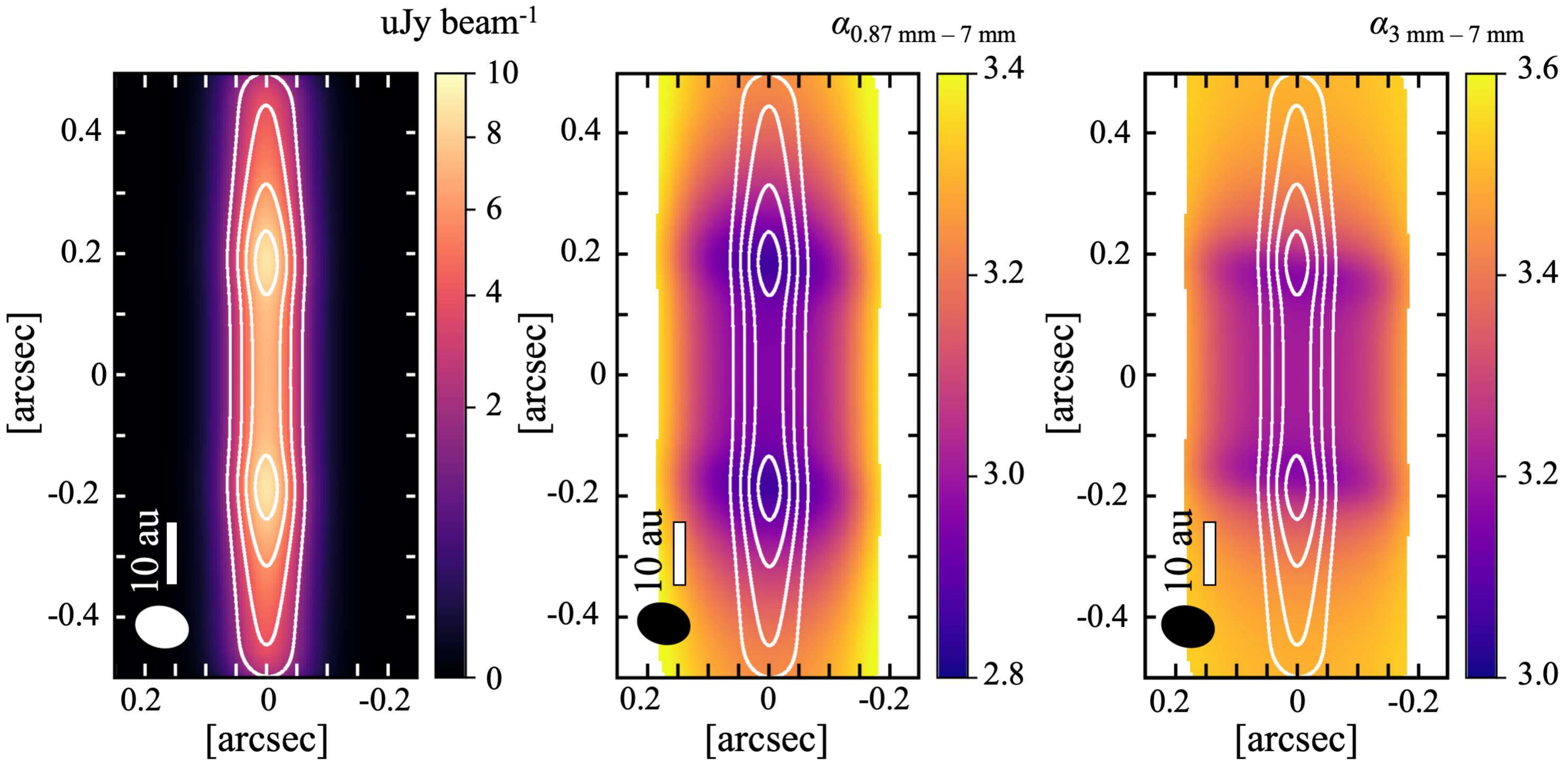}
\caption{The edge-on views of the 7 mm continuum emission (right panel) and spectral index (left panel) maps of the model disk at $t=1.3\times10^4$ yr.
The contours show the continuum intensities of $2,4,6,$ and 8 $\mu$Jy beam $^{-1}$. The beam size is set to be 0.086 arcsec $\times$ 0.067 arcsec with the position angle of 76 degree same as the VLA observations. Note that the absolute values of the intensity and spectral index are different from the L1527 observations because we simply apply the MMSN model in this study.
}
\label{edgeon}
\end{figure}

To compare the L1527 disk with the growth front, we show the dust coagulation model images of the 7 mm continuum emission and spectral index maps of $\alpha_{\rm 0.87 mm - 7 mm}$ and $\alpha_{\rm 3 mm - 7 mm}$, at $t=1.3\times10^4$ yr from edge-on view in Figure \ref{edgeon}.
 The MMSN model with one solar mass protostar is applied for this simulation as the fiducial model.
Note that the face-on view is already shown in Figure \ref{image}.
The model images are smoothed with the VLA beam size of $0\farcs086\times0\farcs067$.
We find that the 7 mm continuum image has double clumps at the growth front radius as similar to the observations because the ring structure has the longest line-of-sight at the edges.
Therefore, we confirm that the observed double clumps are explained by the ring structure.
The substructure is found more clearly in the edge-on disk than the face-on disk with 7 mm continuum emission because the contrast of the surface density is enhanced in the edge-on view.
We note that the intensity and the spectral index values of the model are quite different from the observations because we use the MMSN model in this study.
The comparison of models and observations on intensity and spectral index is only a qualitative discussion.

The spectral index maps, $\alpha_{\rm 0.87 mm - 7 mm}$ and $\alpha_{\rm 3 mm - 7 mm}$, of our model are derived by the intensities between 870 $\mu$m and 7 mm and between 3 mm and 7 mm  wavelengths.
We find that the edge-on disk shows different spectral index pattern from the face-on disk at the growth front.
In the face-on disk, the spectral index peaks at the growth front with $\alpha_{\rm 0.87 mm - 7 mm}\sim4.2$ as shown in Figure \ref{image_alpha}, whereas in the edge-on disk, the spectral index decreases at the growth front with $\alpha_{\rm 0.87 mm - 7 mm}\sim2.8$ and $\alpha_{\rm 3 mm - 7 mm}\sim3.0$ as shown in Figure \ref{edgeon}.
The difference between the edge-on and face-on disk models is the optical depth.
The optical depth at the growth front in the edge-on disk becomes higher than the face-on disk.
The optically thick emission follows black body radiation, which means the spectral index $\alpha=2$.
By taking into account the beam dilution of the VLA observations, the spectral index becomes larger than 2 but lower than that of the optically thin case.
The self-scattering of dust grains \citep{kat15} may also affect the intensity of the dust thermal emission and spectral index at millimeter wavelengths \citep[e.g.,][]{soo17,ued20}.
\citet{liu19} and \citet{zhu19} showed that the spectral index decreases by scattering if emission is optically thick because the scattering of (sub)millimeter-sized dust grains makes the (sub)millimeter thermal emission fainter.
As a result, the spectral index becomes flatter or steeper than the case without scattering effect.
Therefore, the spectral index of the growth front can be changed by the optical depth and effect of the scattering.
However, Figure \ref{edgeon} shows that the increase of the spectral index from inside  to outside the growth front is still remained because dust grains are large/small enough to ignore the scattering effect.

These models assume the same scale height for different sized grains. We consider that the scale height would be less affected at least current observations because all of the observations were not able to resolve the scale height structure. The emission between 870 $\mu$m and 7 mm will be mostly from the disk midplane. 

\begin{figure}[htbp]
\includegraphics[width=8.cm,bb=0 0 2383 1537]{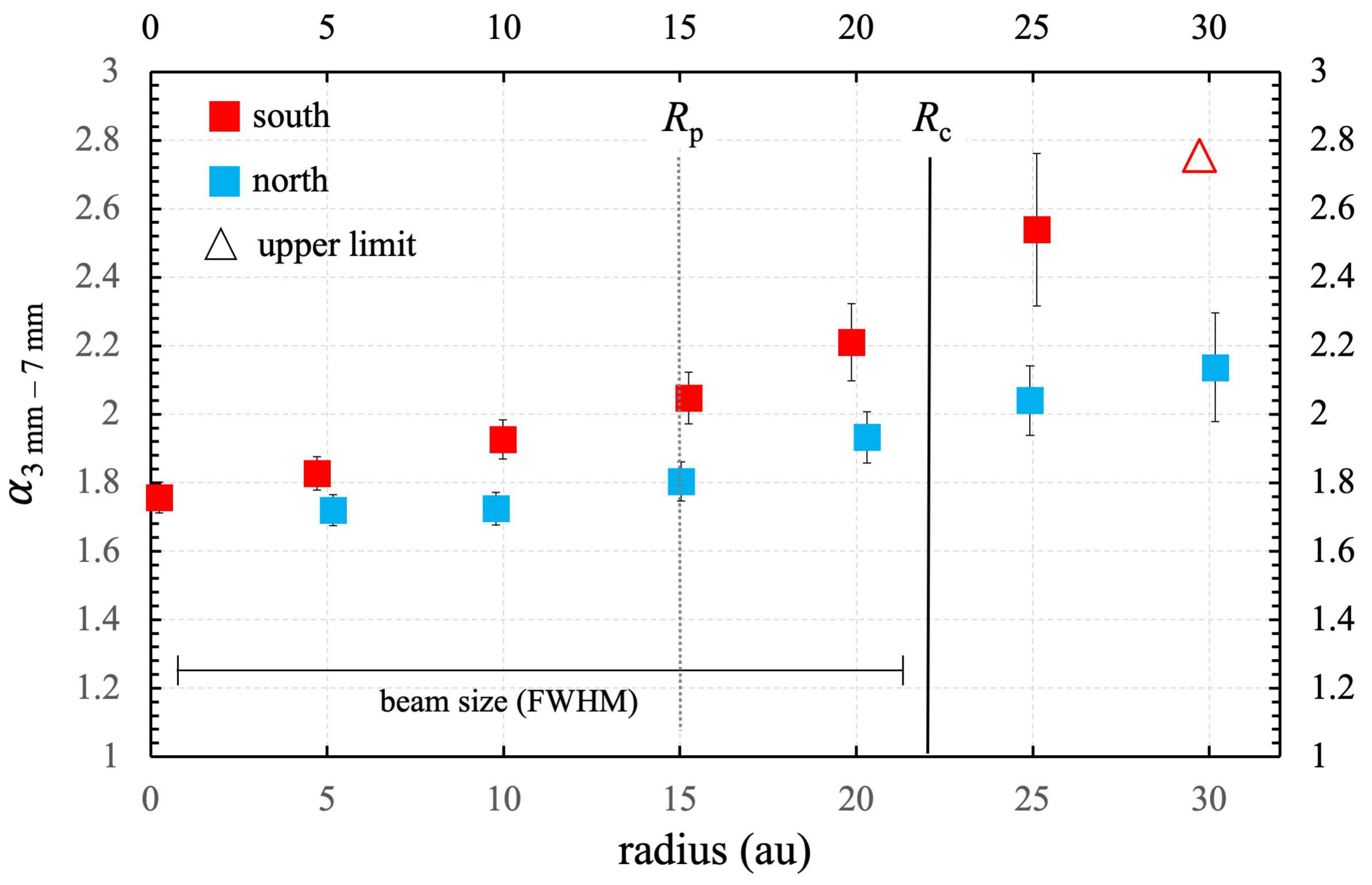}
\caption{The radial profile of the spectral index $\alpha$. The red squares indicate the spectral index along the north direction, while the blue squares indicate that along the south direction.
The error bars represent $\pm1\sigma$. The upper limit of the spectral index is derived by 3$\sigma$ of the VLA continuum emission. The ring position and growth front are shown by the vertical dash dots and black line, respectively. 
}
\label{alpha_radial}
\end{figure}

Figure \ref{alpha_radial} shows the radial profile of the observed spectral index $\alpha_{\rm 3 mm - 7 mm}$ overlaid with the ring position ($R_{\rm p}=15$ au) as the grey dots and with the growth front ($R_{\rm p}=22$ au) as the black line.
The red squares indicate the spectral index along the north direction, while the blue squares indicate that along the south direction.
The upper limit of the spectral index is also derived by using 3$\sigma$ of the VLA continuum emission.

We find that the spectral index becomes lower than 2 in the inner radius even though optically thick emission follows black body radiation with the spectral index $\alpha=2$.
Therefore, the VLA Q Band emission in the inner region will be affected by free-free emission from the protostar and may also be affected by the dust scattering.
Even though there is the contamination of the free-free emission, the spectral index $\alpha_{\rm 3 mm - 7 mm}$ seems to increase outside the ring.
This trend still remains even after extracting the free–free contamination \citep{nak20}.
The increase in the spectral index along the radius indicates that the dust grains are smaller outside the radius.
On the other hand, $\alpha\sim2$ in the inner radius indicates the dust grains have already grown and/or dust continuum emission is optically thick.

The behavior of the spectral index across the ring position is consistent with the idea of the growth front even though the absolute value of the spectral index is different from the model.
The coagulation and grain growth proceed inside the growth front, resulting in the lower spectral index.
On the other hand, the dust grains outside the growth front are not evolved yet.
Therefore, the spectral index is higher in the outer radius than the inner radius.
However, the spatial resolution is not enough to identify the sharp transition of the spectral spectral index across the growth front.

We note that at least the 870 $\mu$m and 3 mm dust continuum emission would be optically thick (this might be also the case for 7 mm dust continuum, in particular for the clump peaks). Therefore, it is difficult to conclude that the dust grains have already grown inside the growth front because the low spectral index can also be explained by high optical depth.
Further observations with high spatial resolution and longer wavelength will allow us to measure the spectral index in more detail.

\subsection{A Caveat on ring formation due to the growth front}

Even though we show that the growth front is consistent with the observed ring location, we point out inconsistency between our model and observations.

The growth front can only explain a single dust ring even though some of the disks have multiple ring structures.
Even in the Class I source of GY 91, three dust rings are observed \citep{she18}. 
As shown in Figure \ref{fig2}, the growth front mainly coincides with the outermost ring.
Therefore, we need additional scenarios to explain entire ring formation(s).
However, recent VLA observations show that  substructures in protostellar disks are dominated by a single bright ring \citep{she20}.

One way to distinguish the mechanisms of the ring formation would be the spectral index $\alpha$ as shown in Figure \ref{image_alpha}.
If ring structures are  formed by a pressure bump due to a presence of planets, dust grains become larger in the ring positions.
On the other hand, the growth front will change the dust grain size inside and outside the ring.
The grain size would be larger on the inside of the ring than on the outside of the ring and ring position.

\section{Discussion and Summary} \label{sec:conclusion}

The location of the growth front ($R_{\rm c}$) is estimated by equation (\ref{eq:front}).
We recall that $R_{\rm c}$ is independent of the dust fluffiness, the disk mass, temperature, or the strength of turbulence.
Therefore, the growth front could be universally observed in various protostellar disks even though we show  the dust surface density of the MMSN model in this study.
Even in the high accretion stage for young protostellar disks, the dust coagulation model is applicable by regarding the high accretion as high turbulence parameter of $\alpha_{\rm D}$ value.

We roughly estimate the occurrence rate of the growth front for observations. 
The growth front will be difficult to be observed if the disk and growth front are small. Therefore, if a disk is younger than $10^3$ yr, the growth front cannot be observed with a spatial resolution of $\sim$ a few au such as ALMA observations. Furthermore, if a disk is more evolved than $\sim3\times10^5$ yr, the growth front is beyond the disk size of 100 au and cannot be observed. Thus, the growth front can only be observable if a disk age is between $\sim 10^3$ and a few $\sim10^5$ yr. By taking into account the lifetime of PPDs of a few Myr \citep[e.g.,][]{hai01}, the occurrence rate will be $\sim$ a few \%. This may be helpful for statistical studies for survey observations in future even though there are some uncertainties for this estimate such as disk size and lifetime. 
Note that this estimate is similar to that of Class 0/I \citep[e.g.,][]{eva09}.
If targets are only limited to Class 0/I sources, the occurrence rate will become much higher.

By comparing with the observations, we found that the ring positions in the YSOs with an age of $\lesssim1$ Myr are consistent with the growth font. Therefore, we propose the growth front to create the ring structure in particular for early stage of the disk evolution such as Class 0 and I sources.
These results indicate that the grain growth via the coagulation occurs quickly and create the ring structure which is observed in a various disks.
For disks with high accretion rates such as Class 0 and I sources, it is difficult to create a situation that suppress radial motion by a ring. 
The growth front scenario is preferred to explain the ring formation and high accretion rate, simultaneously.

In contrast, the growth front extends much larger than the observed ring positions and even disk radii in the protoplanetary disks after 1 Myr.
The observed rings in such late stage disks would be caused by different mechanisms rather than the growth front.
Since the growth front have already swept the entire disk, it may be possible to create the ring/gap structures due to a presence of planets, snow lines of molecules, dust sintering and other mechanisms in those protoplanetary disks after 1 Myr.
By taking into account the results that sufficient mass remains for planet formation in Class 0/I but disappears in Class II \citep[e.g.,][]{man18,tob20}, the planet formation may begin after the passage of the growth front.

The existence of the growth front can be found by changing of dust size across the growth font.
Analysis of the dust spectral index is an important tool for constraining the dust particle sizes in the disk.  We have investigated the disk around the L1527 protostar by using the ALMA Band 3 and VLA Q Band continuum emission as a case study.
The behavior of the spectral index $\alpha$ around the ring position is consistent with the idea of the growth front because $\alpha$ increases outside the growth front even though the absolute value is different from the edge-on disk model.
We note that the comparison of models and observations on intensity and spectral index is only a qualitative discussion.
We suggest that the emission at least ALMA Band 3 may be optically thick with insufficient spatial resolution, and Q Band emission will be affected by the free-free emission from the protostar.
Future observations with high spatial resolution and longer wavelength toward various young disks will allow us to measure the spectral index in more detail.

\acknowledgments
We gratefully appreciate the comments from the anonymous referee that significantly improved this article.
This paper makes use of the following ALMA data: ADS/JAO.ALMA\#2017.1.00509.S. ALMA is a partnership of ESO (representing its member states), NSF (USA) and NINS (Japan), together with NRC (Canada), MOST and ASIAA (Taiwan), and KASI (Republic of Korea), in cooperation with the Republic of Chile. The Joint ALMA Observatory is operated by ESO, AUI/NRAO and NAOJ.
The National Radio Astronomy Observatory is a facility of the National Science Foundation operated under cooperative agreement by Associated Universities, Inc.
This work was supported by JSPS KAKENHI Grant Numbers, 20K14533, 20H00182, 20H04612, 18H05436, 18H05438, 17H01105, 17K05632, 17H01103, and 19K23469.
R.N. and Y.Z. are supported by the Special Postdoctoral Researchers (SPDR) Program at RIKEN.
Data analysis was in part carried out on common use data analysis computer system at the Astronomy Data Center, ADC, of the National Astronomical Observatory of Japan.

\facilities{ALMA, VLA}

\software{RADMC-3D \citep{dul12}
          }

%% This command is needed to show the entire author+affiliation list when
%% the collaboration and author truncation commands are used.  It has to
%% go at the end of the manuscript.
%\allauthors

%% Include this line if you are using the \added, \replaced, \deleted
%% commands to see a summary list of all changes at the end of the article.
%\listofchanges

\appendix\section{Equations of relative velocities induced by Brownian motion, radial drift, azimuthal drift, vertial settling, and turbulence }\label{turbulence}

The Brownian component is given by 
\begin{equation}
\Delta v_{\rm B} = \sqrt{\frac{8 (m_1+m_2)\kB T}{\pi m_1m_2}},
\label{eq:brown}
\end{equation}
where $m_1$ and $m_2$ are the masses of the colliding aggregates and we evaluate this by setting  $m_1 = m_2 = m_{\rm p}$.

The differential drift velocities $\Delta v_{\rm r}$, $\Delta v_{\rm \phi}$, and $\Delta v_{\rm z}$ are functions of the Stokes numbers ${\rm St}_1$ and ${\rm St}_2$ of the colliding pair, which are given by $\Delta v_r= |v_r({\rm St_1}) - v_r({\rm St_2})|$, $\Delta v_{\rm \phi}=|v_{\rm \phi}({\rm St_1}) - v_{\rm \phi}({\rm St_2})|$, and $\Delta v_{\rm z}=|v_{\rm z}({\rm St_1}) - v_{\rm z}({\rm St_2})|$, respectively\citep[e.g.,][]{orm07,oku12}. Since the real size distribution has a finite width, the choice of ${\rm St}_1={\rm St}_2={\rm St}(m_{\rm p})$ results in a significant underestimation of the particle velocities.
\citet{sat16} indicated that ${\rm St}_1={\rm St}_2 /2$ best reproduces the results of a coagulation simulation that treats the fill size distribution. We adopt this relation between ${\rm St}_1$ and ${\rm St}_2$ throughout the paper.
Then, the drift velocities are expressed as
\begin{equation}
\Delta v_{\rm r} =   \Big(\frac{ {\rm 2St_1} } {1 + \rm St_1^2} -   \frac{ {\rm 2St_2} } {  1 + \rm St_2^2}\Big) \eta v_{\rm K},
\label{eq:drift1}
\end{equation}
\begin{equation}
\Delta v_{\rm \phi} = \Big( \frac{1} {1 + \rm St_1^2} - \frac{1}{1 + \rm St_2^2} \Big)  \eta v_{\rm K} ,
\label{eq:drift2}
\end{equation}
\begin{equation}
\Delta v_{\rm z} = \Big( \frac{ \rm St_1}{1 + {\rm St_1}} - \frac{\rm St_2}{1 + {\rm St_2}} \Big) \frac{ h_{\rm d1} h_{\rm d2} }{ \sqrt{\pi (h_{\rm d1}^2 + h_{\rm d2}^2) } } v_{\rm K}.
\label{eq:drift3}
\end{equation}

The turbulent velocity $\Delta v_{\rm t}$ has expressions (see Eqs.~(17) and (18) of \citealt{orm07}) depending on the St and Re$_{\rm t}$ numbers, where ${\rm Re}_{\rm t} =  \alpha_{\rm D} h_{\rm g} c_{\rm s} /\nu_{\rm mol}$ is the turbulent Reynolds number and $\nu_{\rm mol} = v_{\rm th}{\lambda}_{\rm mfp}/2$ is the molecular viscosity. Note that $v_{\rm th} = \sqrt{8\kB T/{\pi} m_{\rm g}}$ and ${\lambda}_{\rm mfp}$ are  the thermal velocity and mean free path of gas particles, respectively.

\begin{equation}
	\Delta v_{\rm t}^2 = \Delta v_{\rm I}^2 +\Delta v_{\rm II}^2,  
	\label{eq:vcolt}
\end{equation}
where
\begin{eqnarray}
	\Delta v_{\rm I}^2 &=& \alpha_{\rm D} c_{\rm s}^2
\frac{\rm St_1- St_2}{\rm St_1 + St_2}\times
\left\{
\left(\frac{\rm St_1^2}{\rm St_{12}^* + St_1}- \frac{\rm St_1^2}{\rm 1+ St_1} \right)   - \left(\frac{\rm St_2^2}{\rm St_{12}^* + St_2}- \frac{\rm St_2^2}{\rm 1+ St_2} \right)  \right\},\label{eq:vcolt1}
	\\
	\Delta v_{\rm II}^2,  &=& \alpha_{\rm D} c_{\rm s}^2
 \Big( ({\rm St_{12}^* - Re_t^{-1/2}}) + \frac{\rm St_1^2}{\rm St_1 + St_{12}^*} - \frac{\rm St_1^2}{\rm St_1 + Re_t^{-1/2} } +  ({\rm St_{12}^* - Re_t^{-1/2} }) + \frac{\rm St_2^2}{\rm St_2 + St_{12}^* } - \frac{\rm St_2^2}{\rm St_2 + Re_t^{-1/2}} \Big)\label{eq:vcolt2}
\end{eqnarray}
and ${\rm St_{12}^*} = {\rm min}(1.6 {\rm St_1},1)$. It should be noted that $\Delta v_{\rm II}^2 = 0$ if ${\rm St_{12}^*} < {\rm Re_t}^{-1/2}$.

\end{document}